\documentstyle[12pt]{article}
\begin{document}

\centerline{\bf (2+0)-DIMENSIONAL INTEGRABLE EQUATIONS AND }
\centerline{\bf EXACT SOLUTIONS}
\vskip1cm
\centerline{E.Sh. Gutshabash{\footnote {e-mail: gutshab@EG2097.spb.edu}},
 V.D. Lipovskii, S.S. Nikulichev }
\vskip0.3cm
\centerline{Institute Research for Physics, St. Petersburg State University,}
\centerline{St.Petersburg, Russia}

\vskip3cm
\centerline{\bf ABSTRACT}
\vskip0.3cm
\parbox[t]{12.7cm}
    {We propose a nonlinear $\sigma$-model in a curved space as a general
integrable elliptic model. We construct its exact solutions and obtain
energy estimates near the critical point. We consider the Pohlmeyer
transformation in Euclidean space and investigate the gauge equivalence
conditions for abroad class of elliptic equations. We develop the
inverse scattering transform method for the $\sinh$-Gordon equation and
evaluate its exact and asymptotic solutions}

\newpage

\centerline{\bf {1. INTODUCTION.}}
\vskip1cm

This work is devoted to studying a broad class of problems
involving two-dimensional nonlinear integrable elliptic equations.
These equations, which describe stationary processes in nonlinear
media, usually,  govern the spatial distributions of physical fields
whose value are fixed at the boundary of the domain under
consideration. The resulting boundary-value problems have a
characteristic feature: as the "observer" penetrates deeper into
the medium (which means considering the asymptotic behavior as
$r^2=x^2+y^2\to\infty$), the field in question become weak, and
the equations become the classical linear equations of
mathematical physics. For this reason, the inverse scattering
transform method (ISTM) applied to the elliptic equations (as well
as the hyperbolic ones) is a nonlinear analogue of the Fourier
transform method, a tool often applied to linear problems. This
analogy allows one, in particular, to interpret the
discrete-spectrum solutions of nonlinear elliptic equations as
"solitonlike" excitations (vortices, or defects) in nonlinear
condensed media. These excitations possess some properties of
solitons and also some special properties of their own.

In [1-4], the boundary-value problems on the half-plane were
posed for two-dimensional elliptic equations, the direct and the
inverse scattering problems were investigated for the
corresponding operators of the associated linear problems, and the
exact solutions, the conservation laws, and the trace identities
were also found. In addition, the boundary-value problems for the
elliptic versions of the two-dimensional Heisenberg ferromagnet
(of the $O(3)\:\sigma$-model) were shown to be gauge equivalent
to the $\sinh$-Gordon equation [4]-[5]. These results have thus
allowed us to extend the methods used with the hyperbolic
equations to the elliptic case, which has numerous important
physical applications.

At the same time, a currently important problem consists in
constructing more-realistic models that would still be integrable
on the one hand and would account for various features of the
physical systems on the other hand. Solving one such problem was
begun in  [6], where a $\sigma$-model in a curved space was
proposed (a magnet with a variable nominal magnetization). We show
in what follows that this model (more precisely, its integrable
version) is the most general of the presently known physical
models that are described by elliptic equations and possess Lax
representations.

In this paper, which is an expanded version of [6, 1] and a
revised version of [7], our aim is to obtain exact solutions of
this model, to investigate its properties, and also to find a chain
of transitions leading from this model to those that have been  known
and investigated before. For one such model
($\sinh$-Gordon), we use the ISTM to construct exact solutions and
give their physical interpretation.

The paper is organized as follows. In Sec.2, we consider a
$\sigma$-model in a curved space, find its exact solutions, and
investigate some thermodynamic properties. In Sec.3, we propose a
Euclidean version of the Pohlmeyer transformation and investigate
the question of the gauge equivalence (GE) of a broad family of
boundary-value problems. \sloppy In Sec.4, we develop the ISTM for
a model of charged particles on a half-plane, find a series of its
exact solutions, and also find the conservation laws and trace
identities.

\vskip2cm
\centerline{\bf {2. NONLINEAR $\sigma$-MODEL IN CURVED SPACE.}}
\vskip2cm

As is well known, the condition that the absolute value of the
magnetic moment be constant is an essential assumption in the
phenomenological description of weakly excited states of magnets.
With this assumption, the evolution of system states amounts to
the rotation of the magnetization vector. A similar statement is
true for magnet models described by the Landau-Lifshits equation.
Such models provide a physically adequate picture far from the
Curie point. For moderate temperatures, however, one can expect a
spatial variation of the absolute (nominal) magnetization of the
sample material. This hypothesis allows one to investigate the
magnet near the critical point (but outside the fluctuational
region) using the phenomenological approach and to describe a
smooth spatial transition to the paramagnetic phase.

We consider the two-dimensional isotropic Heisenberg ferromagnet,
i.e., the two-dimensional stationary Landau-Lifshits equation

$$
{\bf M} \wedge \triangle {\bf M}=0,\;\;\;{\bf M}^2=\alpha({\bf r}),\eqno(2.1)
$$
where ${\bf M}={\bf M}(x,y)=(M_1,M_2,M_3),\; {\bf r}=(x,y)\in
R^2_{+},\;$ and $\alpha({\bf r})$ is the square of the nominal
magnetization, which is an arbitrary function of ${\bf r}$, in
general.

The equation
$$
\alpha\triangle {\bf M}+({\bf M}^2_x+{\bf M}^2_y){\bf M}=
\frac {1}{2}{\bf M}\triangle\alpha\
\eqno(2.2)
$$
is easily obtained from (2.1). We set $$
M=\sum_{i=1}M_i\sigma_i,$$ where $\sigma_i,\; i=1,2,3,$ are the
Pauli matrices. In terms of the matrix $M$, Eq. (2.2) is

$$
(\alpha M_zM^{-1})_{\bar z}+(\alpha M_{\bar z}M^{-1})_z=\alpha_{z\bar z}I,\
\eqno(2.3)
$$
where $z=x+iy$ and $I$ is the $2\times 2$ unit matrix.

Equation (2.2) (or the equivalent equation (2.3)) describes an
inhomogeneous magnet, a generalization of the two-dimensional
version of the stationary Landau-Lifshits magnet. Since the
question of the integrability of these equations remains
unanswered in the general case of an arbitrary function
$\alpha({\bf r})$), we consider the simplest version, where Eq.
(2.3) is integrable, obtained by setting
$$
4\triangle \alpha=\alpha_{z\bar z}=0.\eqno(2.4)
$$
Then (2.3) implies that a two-dimensional isotropic magnet whose
squared saturation moment is a harmonic function of the
coordinates is described by a nonlinear $\sigma$-model in a curved
space.

We note an interesting analogy with gravity theory. Equation (2.3)
under condition (2.4) has the same form as the gravity equations
with two commuting Killing vectors for the part of the metric $g$
containing the off-diagonal terms [8-11]. The difference is that
we have $M=M^{*}$ (where $*$ denotes the Hermitian conjugation)
and
$
\det M=-\alpha $ here and $g=g^T$ and
$\det g=-\alpha^2$ in the gravity case.

We set
$$
 S=M/\sqrt{\alpha},\;\;\;  S=\sum_{i=1}^3S_i\sigma_i.
$$
Therefore, ${\bf S}^2=1$, i.e. ${\bf S}$ is the normalized
magnetization vector. In accordance with (2.3) and (2.4), we now
have
$$ (\alpha S_zS)_{\bar z}+(\alpha S_{\bar z}S)_z=0,\eqno(2.5) $$
which is the equation of a nonlinear  $O(3)\: \sigma$-model in a
curved space. If $\alpha=1,$ Eq. (2.5) becomes the equation of the
standard $O(3)\: \sigma$-model (the two-dimensional isotropic
Heisenberg ferromagnet), for which the boundary-value problem was
solved in [3]-[4]. We also note that the above-mentioned case of
gravity reduces to a similar $O(2,1)\: \sigma$-model in a curved
space and that dimensional reduction of the axially symmetrical
self-duality equations in the Yang-Mills theory also leads to a
similar model. The analogue of $S$ is then given by $g^{*}g$,
where $g\in SL(N,R)$ [12].

We first consider some simple automodel solutions of (2.5). We
take the ansatz
$$
S=\left(\matrix{\cos \chi& \sin \chi\exp (-i\Phi)\cr
\sin \chi\exp ( i\Phi)&-\cos \chi\cr}\right),\eqno(2.6)
$$
where $\chi=\chi(\alpha,\beta),\Phi=\Phi (\alpha,\beta)$ is a
real-value function, and $\beta$ is the harmonic function that is
conjugate to $\alpha$ in the sense of the Cauchy-Riemann
conditions. Substituting (2.6) in (2.5) produces the system of
equations

$$
(\Phi_{\alpha\alpha}+\alpha ^{-1}\Phi_{\alpha}+\Phi_{\beta\beta})\sin \chi+\
2(\Phi_{\alpha}\chi_\alpha+\Phi_{\beta}\chi_{\beta})\cos \chi=0,
$$
$$
{}\eqno(2.7)
$$
$$
2(\chi_{\alpha\alpha}+\alpha ^{-1}\chi_{\alpha}+\chi_{\beta\beta})=\
(\Phi^2_{\alpha}+\Phi^2_{\beta})\sin 2\chi
$$
For $\chi=0,$ we have $S=\sigma_3$. For $\chi=\pi/2,$ the function
$\Phi$ satisfies the Laplace equation; it then follows that
$$
\Phi(\alpha,\beta)=\int dR(s)Z_0(is\alpha)e^{is\beta} ,
$$
where $Z_0(x)$ is the Bessel function ($J_0$ or $N_0$) and $dR(s)$
is a spectral measure. Finally, for $\Phi=k\beta$ with a free
parameter $k,$ the function $\chi=\chi(\alpha)$ satisfies the
equation
$$
\chi_{\alpha\alpha}+\alpha^{-1}\chi_{\alpha}=\frac {k^2}{2}\sin 2\chi ,
$$
which reduces to the third Painlev\'e equation.

System (2.7) is analogous to the equations that arise in gravity
theory [13]; however, one of those equations resembles the
Liouville equation rather than the sine-Gordon equation as is the
case here.

Equation (2.5) is the compatibility condition for the linear
$2\times 2$ matrix system
$$
(\partial_z-\frac {\alpha S_zS}{\varrho+\alpha})\Psi=0, \;\;\;
(\partial_{\bar z}+\frac {\alpha S_{\bar z}S}{\varrho-\alpha})\Psi=0, \
\eqno(2.8)
$$
where
$$
\varrho=\varrho (\alpha,\beta,u)=i\beta-u+\sqrt{(u-\gamma)\
(u+\bar {\gamma})},\;\gamma(z)=\alpha+i\beta,\eqno(2.9)
$$
$\gamma$ is the function that is analytic within the sample, and
$u\in C$ is a "hidden" spectral parameter that does not depend on
the coordinate.
Thus, Equation (2.5) can be solved by a general version of ISTM.
This version, however, has not yet been fully described in view of the
technical complications (some progress was recently made in [14] for
the gravity theory equations). Therefore, we use the dressing method
for deriving exact solutions. Because the strategy in this case is
essentially the same as in [15], we give only parts of the calculations
that allow us to obtain the final answer.

To system (2.8), we add the reduction conditions and the formula for
reconstructing the potential (magnetization):
$$
\Psi^{-1}(u)=\Psi^*{(-\bar u)},\;\; S\Psi(u)=\Psi(\tau(u))J,\;\;
S=\Psi(\infty^{+})C,
$$
$$ \eqno(2.10)
$$
$$
JJ^{*}=J^2=I,\;\; CC^{*}=I,\;\; J=C\Psi(\infty^{-}),
$$
where $\tau$ is the
transposition of the sheets of the Riemann surface $\Gamma$ of the
square root entering (2.9) and the signs + and - pertain to the upper and
the lower sheet, with the upper sheet being fixed by the condition that
$\varrho\to 0$ as $u\to \infty ^{+}$. Further, the $2\times
2$ matrices $C$ and $J$ are introduced to select different
versions of the background solutions (the classical vacuum). In
addition, the involution conditions
$$ \varrho(u)=-\bar \varrho(-\bar u),\;\
\varrho(\tau(u)=\alpha^2/\varrho(u)\eqno(2.11) $$
hold. Relations (2.9)-(2.11) allow us to write the answer. We seek the
solution of (2.8) in the form
$$
\Psi(u)=\chi(u)\Psi^{0}(u),\eqno(2.12) $$
where $\Psi^{0}(u)$ is the bare solution of (2.8) corresponding to the
vacuum $S^0$ and the matrix $\chi(u)$ is to be determined. We further assume
that the matrices $\chi(u)$ and
$\chi^{-1}(u)$ are meromorphic functions on the Riemann surface $\Gamma$
and have a finite number of poles (those of the $\chi$ matrix occurring at
the points $u=v_i$),
$$ \chi(u)=I+\sum_i \frac {|m_i>\otimes
<q_i|}{\varrho(u)-\mu_i},\;\; \chi^{-1}(u)=I+\sum_i \frac {|p_i>\otimes
<l_i|}{\varrho(u)-\nu_i},\ \eqno(2.13) $$ where $\mu_i=\varrho(v_i)$,
$\nu_i$ are some complex numbers specified below, and the standard
Dirac notation for vectors is used.

Obviously, if $\chi$ has a pole at $u=v_i$, then it has a pole on the
second sheet of $\Gamma$ at $u=\tilde
v_i=\tau(v_i)$. It follows from the condition
$\chi(u)\chi^{-1}\ (u)=I$ and Eq. (2.13) that
$$
|m_i>=\sum_j |p_j>(N^{-1})_{ji},\;\; <l_i|=-\sum_j (N^{-1})_{ij}<q_j|,\
\;N_{ij}=\frac {<q_i|p_j>}{\mu_i-\nu_j}. $$
Equations (2.8), (2.12)-(2.13) allow us to find
$$ <q_i|=<d_i|(\Psi^0(v_i))^{-1},\;\;\;|p_i>=\Psi^0(-\bar v_i)|c_i>, $$
where $<d_i|$ and $|c_i>$ are arbitrary bra and ket vectors. The first of
the reductions in (2.10) gives $<q_i|=|p_i>^*$ and $\nu_i=-\bar \mu_i$,
and the second one implies that $<c_i|J=<\tilde
c|$.  Here, the vector $<c_i|$ corresponds to the pole
contribution coming from $u=v_i$ and
$<\tilde c_i|$ to the contribution of $u=\tilde v_i=\tau(v_i).$

The relations obtained allow us to write the "$N$-soliton"
solution of (2.5); for simplicity, we limit ourselves to the
"one-soliton" solution
$$ S=S_0+\frac {<q|S^0|q>}{D(|\mu|^2+\alpha^2)}\left.\left [S^0|q>\otimes
<q|S_0+\ \frac {|\mu|^2}{\alpha^2}|q>\otimes <q|\right.\right]
- $$ $$ {}\eqno(2.14) $$
$$ -\frac {<q|q>}{\alpha^2D(\mu+\bar \mu)}
\left.\left[\bar \mu|q>\otimes<q|S^0+\mu
S^0|q>\ \otimes <q| \right.\right]. $$
Here, $\mu=\varrho(v)=\mu_R+i\mu_I,\:|q>=\Psi^0(-\bar
v)|c>,\:<q|=|q>^*,\:|c>\in C^2$ is an arbitrary constant vector, and
$$ D=\frac {|\mu|^2}{\alpha^2}\left.\left[\frac
{<q|q>^2}{(\mu+\bar \mu)^2}-\ \frac
{\alpha^2<q|S^0|q>^2}{(|\mu|^2+\alpha^2)^2}\right.\right]. $$
Equation (2.14) allows one to construct the "$N+1$-soliton" solution
starting with an "$N$-soliton" back-ground.

We next analyze the explicit form of the solution obtained. The
simplest choice for the background, $S^0=\sigma_3,$ corresponds to
$C=I,\;J=\sigma_3,\;\Psi^0=\sigma_3$.  Setting $<c|=(1,\bar c)$
(without losing generality), we have from (2.14) that
$$
S_{+}=S_1+iS_2=-\frac {c}{\alpha^2D}
\left.\left[(1-|c|^2)\frac {|\mu|^2-\alpha^2}{\
|\mu|^2+\alpha^2}+(1+|c|^2)\frac {\mu-\bar \mu}{\mu+\bar \mu}\right.\right],\
S_3=1-\frac {2|c|^2}{\alpha^2D},
{}\eqno(2.15)
$$
where
$$
\;
D=\frac {|\mu|^2}{\alpha^2}\left.\left[\frac {(1+|c|^2)^2}{(\mu+\bar \mu)^2}-\
\frac {\alpha^2(1-|c|^2)^2}{(|\mu|^2+\alpha^2)^2}\right.\right].
$$
When $c=-i\exp (2i\Phi)$, Eqs. (2.15) are simplified
considerably and become
$$
S_{+}=-\sin 2\theta e^{2i\Phi},\;\;\;
S_3=-\cos 2\theta,\eqno(2.16)
$$
where $\theta=\arg \mu$. This reproduces ansatz (2.6) with
$\Phi=const$ and $\chi=\pi+2\theta$. Moreover, it can be shown that
$\chi$ satisfies the second equation in (2.7) with a
vanishing right-hand side. For a purely imaginary $v$, it follows from
(2.11) and (2.16) that the excitation over the $S_3^0=1$
vacuum is absent. This is true also in the general case
described by Eq. (2.14).  We also note that solutions (2.15) and (2.16)
have no nontrivial analogues in the case of a constant saturation
moment.

Another possible choice of the bare solution,
$S^0=\exp (-ik\beta\sigma_3)\;\sigma_1$, where $k$ is a real constant,
leads to the equations $C=J=\sigma_1$.  In this case, system (2.8)
can be explicitly integrated as
$$ \Psi^0(\alpha,\beta,u)=e^{k\sigma_3(\varrho(\alpha,\beta,u)-2i\beta)/2},$$
where $\varrho$
is defined in (2.9). Again setting $<c|=(1,\bar c)$, we can
generalize the formulas describing magneticovortex excitations [16]
to the inhomogeneous case under consideration:

$$
S_{+}=e^{ik\beta}+\frac {4|c|^2|\mu|}{\alpha D}e^{ik\beta}\left.\left[\frac {\
\cos(\eta+i\sigma)\cos \eta}{|\mu|^2+\alpha^2}-\frac {\cosh
{\xi}}{2\alpha\mu_R}\ \cosh {(\xi-i\theta)}\right.\right],
$$
$$
{}\eqno(2.17)
$$
$$
S_3=-\frac {2|c|^2}{\alpha^2D}
\left.\left[\frac {|\mu|^2-\alpha^2}{|\mu|^2+\alpha^2}\
\sinh \xi\cos \eta-i\frac {\mu-\bar \mu}{\mu+\bar \mu}\cosh \xi\sin \eta
\right.\right],
$$
where
$$
D=\frac {4|\mu|^2|c|^2}{4\mu_R^2}\left[ \frac {\cosh ^2 \xi}{4\mu_R^2}-\
\frac {\alpha^2 \cos^2 \eta}{(|\mu|^2+\alpha^2)^2}\right] .
$$
In these formulas, $\xi=k\mu_R+\ln |c|$, $\eta=k(\mu_I-\beta)-\arg
c$, $\theta=\arg \mu$, $\sigma=\ \ln (|\mu|/\alpha)$. The formulas
for $S_{+}$ and $S_3$ give the magnetization distribution
characterized by local variation scale $L^{-1}\sim \ |k\bigtriangledown
{\mu_R}|$ and the wave number $K\sim|k\bigtriangledown\
(\mu_I-\ \beta)|$, which correspond to the spatial
modulation of the magnetic structure of the sample. For $KL\gg 1$,
relations (2.7) describe an inhomogeneous spiral magnet that can
produce multiple harmonics of the fundamental frequencies.

The above "solitons" allow us to estimate the energy functional
evaluated on these solutions. To do so, we assume that the
function $\alpha$
depends on the temperature in addition to the spatial variables:
$\alpha=\alpha({\bf r},T)$.  The exact general form of this
dependence can be found by considering the system on a lattice
[17,18], which is beyond the scope of this paper. We therefore
merely estimate the energy and the heat capacity near the critical
point.

According to the Landau theory [19], the functions $\alpha$ and $\beta$
behave as follows near the Curie point $T_c$:
$$ \alpha (r,T)\approx f_0(r)t,\;\;\;\beta (r,T)\approx
g_0(r)t,\eqno(2.18) $$
where $t=(T_c-T)/T_c$. The second estimate in (2.18)
follows from the first one and
fact that the quantities $\alpha$ and $\beta$ (as well
as $f_0$ and $g_0$) are related by the Cauchy-Riemann conditions.
Taking $\alpha$ and $\beta$ to be small and
$ t \to 0^{+}$, we see from (2.9) that
$$ \mu=-\frac
{\alpha^2}{2v}(1+\frac {i\beta}{v})+O(|\gamma|^4)\eqno(2.19). $$

The system energy is defined as
$$ E=\frac {1}{2}\int_\Omega dxdy
(\bigtriangledown {\bf M})^2=\ \frac {1}{2}\int_\Omega dxdy [\frac
{(\bigtriangledown \alpha)^2}{4\alpha}+\ \alpha(\bigtriangledown {\bf
S})^2],\eqno(2.20) $$
where $\Omega$ is the domain occupied by the sample. This relation can be
rewritten as
$$ E=\int_\Omega d^2r[\alpha Tr(S_zS_{\bar z})+\frac
{\alpha_z\alpha_{\bar z}}{\ 2\alpha}]=E_1+E_0.\eqno(2.21) $$
Using $$ d^2r=dxdy=-\frac {1}{2i}d \gamma \wedge d \bar \gamma
\frac {1}{\gamma_z\ \bar \gamma_{\bar z}},\;\;\; d\alpha \wedge d\beta=-\frac
{1}{2i}d\gamma\wedge d\bar \gamma, $$
we obtain
$$ E_0=\frac {1}{8}\int_{\Omega^{\prime}} \frac {d\alpha d\beta}{\alpha},\;\;
\; E_1=\int_{\Omega^{\prime}} \frac {d\alpha d\beta}{\gamma_z\ \bar
\gamma_{\bar z}}\ Tr(S_zS_{\bar z}), \eqno(2.22) $$
where $\Omega^{\prime}$ is the domain in the plane of the
$(\alpha,\beta)$ variables obtained via the conformal mapping $\gamma(z)$
of $\Omega$ from the $(x,y)$ plane.  It follows from (2.22) that
$$ E_0=\frac {1}{8}(\beta_2-\beta_1)\ln {\frac
{\alpha_2}{\alpha_1}\sim {\ (T_c-T)\ln (T_c-T)}}, \eqno(2.23) $$
from which, in particular, we can find the contribution to the heat
capacity as
$$
c_0=\frac {\partial E}{\partial T}\sim {\ln [e(T_c-T)]}. \eqno(2.24) $$

Estimating $E_1$ and $c_1=\partial {E_1}/\partial T$ requires
knowing the explicit form of the solution for $S$.  We first
consider the simplest solution (2.16). We then find
$$ E_1=8\int_{\Omega^{\prime}} \frac
{d\alpha d\beta}{\gamma_z\ \bar {\gamma_{\bar z}}}\
\alpha\theta_z\theta_{\bar z}=\ 8\int_{\Omega^{\prime}} d\alpha
d\beta\alpha\theta_{\gamma} \theta_{\bar \gamma},\eqno(2.25) $$
with $\theta_{\gamma}$ and $ \theta_{\bar {\gamma}}$ satisfying the
relations
$$ \theta_{\gamma}=\frac {1}{2i}(\ln(\frac {\varrho}{\bar
{\varrho}}))_{\gamma},\;\; \; \theta_{\bar {\gamma}}=\frac {1}{2i}(\ln(\frac
{\varrho}{\bar {\varrho}}))_{\ \bar {\gamma}}\eqno(2.26) $$
Using the deformation method developed in [20], we can show that
$$ \left.\left(\frac{4\alpha}{\mu-\alpha}\right.\right)_{\gamma}=
\frac {1}{\mu-\alpha}+\ \frac
{1}{\mu+\alpha},\;\;\; \left.\left(\frac {4\alpha}
{\mu+\alpha}\right.\right)_{\bar \gamma}=\ \frac
{1}{\mu-\alpha}+\frac {1}{\mu+\alpha},\eqno(2.27) $$
from which, setting $\mu=\varrho$, we obtain
$$ \theta_{\gamma}=\frac {1}{2i}\frac {\varrho+\bar
\varrho}{(\varrho+\alpha)\ (\bar {\varrho}-\alpha)},\;\;\; \theta_{\bar
{\gamma}}=-\ \frac {1}{2i}\frac {\varrho+\bar \varrho}{(\varrho-\alpha)\
(\bar {\varrho}+\alpha)}.\;
$$
Thus, in accordance with (2.25), we have
$$
E_1=2\int_{\Omega^{\prime}} d\alpha d\beta\
\frac {\alpha(\varrho+\bar \varrho)^2}{\
|\varrho^2-\alpha^2|^2}.\eqno(2.28)
$$

Using (2.18) and (2.19) and taking the size of $\Omega^{\prime}$
to be finite, we finally obtain the following relations
satisfied near the Curie point:
$$ E_0\approx t\ln t,\;\; c_0\approx \ln t,\;\; E_1\approx
t^3,\;\; c_1\approx t^2.\ \eqno(2.29) $$
Thus, the background heat capacity satisfies the law characteristic of
the two-dimensional Ising model [17, 18] with the critical index equal
to zero, and the energy and the heat capacity of the excited state
over the vacuum satisfy powerlike laws.

For convenience in calculating the energy of more complicated
solutions, we rewrite Eq. (2.21) as
$$
E=\int_{\Omega^{\prime}} d\alpha d\beta\:[\frac {\alpha}{\gamma_z\ \bar
{\gamma_{\bar z}}}\ (2S_{3z}S_{3\bar z}+|S_{+z}|^2+|S_{+\bar z}|^2)+\frac
{1}{8\alpha}].\eqno(2.30) $$

We consider solution (2.15), setting $|\mu|=\alpha \tan \Phi$ and $\Phi \in
[0,\pi /2)$.  Then,

$$
S_{+}=-\frac {c}{\alpha^2D}[-B\cos 2\Phi+iA\tan \theta],\;\;\;
S_3=1-\frac {2|c|^2}{\alpha^2D},\eqno(2.31)
$$
where
$$
D=\frac {1}{4}\frac {A^2}{\alpha^2\cos ^2\theta}-B^2\sin ^2 \Phi,\;\;\;
D=\bar D,
$$
$\theta=\arg \mu,\: A=1+|c|^2,\: B=2-A$.
Using the deformation method, we obtain
$$
\Phi_z=\frac {\cos^2{\Phi}}{\alpha}\bar {\gamma}_z\left.\left[|\varrho|^2\
\frac {\bar \varrho -\varrho-2\alpha}{(\varrho+\alpha)(\bar
\varrho -\alpha)}-\ \frac {1}{2}\tan \Phi\right.\right], $$
$$ {}\eqno(2.32) $$ $$
\Phi_{\bar z}=\frac {\cos^2{\Phi}}{\alpha}\bar {\gamma_{\bar z}}
\left.\left[|\varrho|^2\
\frac {\bar \varrho -\varrho-2\alpha}{(\varrho-\alpha)(\bar \varrho+
\alpha)}-\
\frac {1}{2}\tan \Phi\right.\right].
$$
Taking (2.31) and (2.32) into account, we obtain
$$
E_0, \; E_1\approx t\ln t,\;\; c_0,\; c_1\approx \ln t, \eqno(2.33)
$$
which means that the background energy and the heat capacity of (2.15) give
the same contributions as the corresponding quantities evaluated
for the states with excitations.

We consider solution (2.17). Using the same notation as in the
previous example,
we write it as
$$
S_{+}=e^{ik\beta}+\frac {2|c|^2}{\alpha^2D}e^{ik\beta}[\cos (\eta+\
i\sigma)\sin 2\Phi\cos \eta-\frac {\cosh \xi}{\cos \theta}\cosh (\xi-\
i\theta)],
$$
$$
S_3=-\frac {2|c|^2}{\alpha^2D}[-\cos 2\Phi\sinh \xi\cos \eta+\tan \theta\
\cosh \xi\sin \eta].
$$
Evaluating the corresponding derivatives and estimating the integrals as
before, we again obtain
$$
E_0,\; E_1\approx t\ln t;\;\;\; c_0,\; c_1\approx \ln t. \eqno(2.34)
$$

The approach described here contains, therefore, three main ingredients:
the integrability of the model, the dressing method, and the
deformation method.
Because of its universality, the approach can be used with other
interesting applications of (integrable) two-dimensional models.

To conclude this section, we show how the transition to the homogeneous
case can be accomplished in the framework of the inhomogeneous magnet
model. We consider system (2.8) and expression (2.9), which contains the
"hidden" spectral parameter $u$. Setting
$\lambda=\varrho/\alpha$ and letting $\alpha$ tend to unity, we obtain
$$ \varrho=i\beta-u+\sqrt {(u-1)(u+1-i\beta)}=\lambda.
\eqno(2.35) $$
Because $\gamma (z,\bar z)=\alpha (z,\bar z)+i\beta
(z,\bar z)$ with $z=x+iy$ is an analytic function, the
Cauchy-Riemann conditions imply that $ \beta\to
{\beta_0}$ as $\alpha\to 1$, where $\beta_0$ is a constant which
we set equal to zero. Then we see from (2.35) that $u=-1/2(\
\lambda+1/\lambda)$, and in the language of spectral parameters, therefore,
the correspondence between the homogeneous and the inhomogeneous case is
given by the "mirrored" Zhukovskii function.

\vskip1cm
\centerline {\bf {3. The Pohlmeyer transformation and gauge equivalence of }}
\centerline{\bf {two-dimensional integrable boundary-value problems}}
\vskip1cm

We consider the nonlinear $O(3) \sigma$-model on the half-plane
 $R^2_{+}=\{(x,y):-\infty<x<+\infty, y \ge 0 \}$:
$$ {\bf S}_{z\bar z}+({\bf S}_z{\bf S}_{\bar z}){\bf S}=0,\eqno(3.1)
$$
where ${\bf S}(x,y)=(S_1,S_2,S_3),\;|{\bf S}|^2=1,$ and ${\bf
S}_z=(1/2)({\bf S}_x-i{\bf S}_y).$ Model (3.1) coincides with the
elliptic version of the ${\bf n}$-field model [21] and is one of the
versions of the chiral field model. We assume that
${\bf S}(x,0)$ and ${\bf S}_y(x,0)$ are given smooth functions such that
(unless otherwise stated)
$${\bf S}(x,0)\to (\cos 2x, \sin 2x, 0), \;\; |x|\to\infty, \eqno(3.2) $$
i.e., ${\bf S}$ belongs to the class of functions describing spiral
structures [22].  Following [23], we also assume that the relations
$$
{\bf S}_z^2={\bf S}_{\bar z}^2=1 ,\eqno(3.3)
$$
hold on $R^2_{+}$. These relations then imply the constraints

$$
{\bf S}_x^2-{\bf S}_y^2=4,\;\;\; {\bf S}_x{\bf S}_y=0,\eqno(3.4)
$$
which are obviously compatible with (3.1).

Setting $f(z,\bar z)=\bf {S}_z\bf {S}_{\bar z}$,
we perform the Pohlmeyer transformation assuming that

$$
{\bf S}_{zz}=c_1{\bf S}+c_2{\bf S}_z+c_3{\bf S}_{\bar z},
$$
where $c_i=c_i(z,\bar z),i=1,3,$ are some complex-valued functions. It
can be easily verified that  $c_1=-1,\; c_2=-{ff_z/(1-f^2)}$, and
$c_3={f_z/(1-f^2)}$. This implies the equation

$$
f_{z \bar z}=1-f^2-\frac {ff_zf_{\bar z}}{1-f^2}.\eqno(3.5)
$$
This equation is integrable, because it is the compatibility condition
of the linear matrix system

$$
\Psi_z=A\Psi,\;\;\;\Psi_{\bar z}=B\Psi,\eqno(3.6)
$$
where $\Psi=\Psi(z,\bar z),\;A=A(z,\bar z),\;B=B(z,\bar z)\in
Mat(2,C)$, and  the matrices $A$ and $B$ have the forms

$$
A=\frac {1}{2}i\lambda \sigma_3+\frac {\partial f}{2\sqrt {f^2-1}}\sigma_1
,\;\; B=\frac {1}{2\lambda}if\sigma_3+\frac {1}{2\lambda}\sqrt
{f^2-1}\sigma_2\ \eqno(3.7) $$
Here, $\lambda \in C$ is the spectral parameter, and $|f|>1$.

In what follows, we use the notion of the "phase" space of a nonlinear
elliptic equation, which can be introduced by a natural analogy with the
hyperbolic case [21]. We do this in the example described by Eqs. (3.1)
and (3.5).

The generating functional of the model given by (3.1) and (3.2) is
$$ F^{({\bf S})}=\int_{-\infty}^{+\infty} dx
\int_0^{\infty} dy\ \left.\left [\frac
{1}{2}({\bf S}_x^2+{\bf S}_y^2)+\frac {\nu_1}{2}({\bf S}^2-1)+\ \nu_2{\bf
S}{\bf S}_y-2\right.\right],\eqno(3.8) $$
where $\nu_j,j=1,2,$ are the Lagrange multipliers. The phase space
$\cal M^{({\bf S})}$ consists of the smooth functions
$S_i(x,0)$ and $S_{iy}(x,0),\:i=1,3$, and the analogues the Hamilton
equations are

$$
{\bf S}_y=\{H^{({\bf S})},{\bf S}\},\;\;\; ({\bf S}_y)_y=\{H^{({\bf S})},{\bf
S}_y\},\ \eqno(3.9) $$
with the "Hamiltonian" of the model given by
$$
H^{({\bf S})}=\int_{-\infty}^{+\infty}dx \left.\left[\frac {1}{2}
({\bf S}_x^2-{\bf S}_y^2)+2 \right.\right ].\eqno(3.10)
$$
The Poisson structures of the phase space is defined by the nonvanishing
fundamental brackets with the constraints involved in (3.8) duly taken
into account [21]:

$$
\{S_{ay}(x),S_b(x^{\prime})\}=[\delta_{ab}-S_a(x)S_b(x)]\delta (x-x'), $$
$$ \{S_{ay}(x),S_{by}(x^{\prime})\}=-[S_{ay}(x)S_b(x)-S_{by}(x)S_a(x)]
\delta (x-x^{\prime}).  $$
For arbitrary smooth functionals $F$ and $G$, we then have
$$ \{F,G\}=\int dx \Bigg\{\left [\frac {\delta F}{\delta
{\bf S}_y}\frac {\delta G}{\ \delta {\bf S}}-\frac {\delta F}{\delta {\bf
S}}\frac {\delta G}{\delta (\delta_y{\bf S})}\ \right]+$$ $$+ \left(\frac
{\delta G}{\delta {\bf S}_y}{\bf S}\right) \left [\left(\frac {\delta
F}{\delta {\bf S}}\ {\bf S}\right)-\left(\frac {\delta F}{\delta {\bf
S}_y}\ {\bf S}_y\right) \right]+\left(\frac {\delta F}{\delta {\bf
S}_y}{\bf S}\right) \left[\left({\delta G\over\delta {\bf S}_y}{\bf
S}_y\right)-\left({\delta G\over \delta {\bf S}}{\bf S}\right)\right]
\Bigg\}.$$

We now consider Eq. (3.5) assuming that
$f(x,0)$ and $f_y(x,0)$ are given smooth real-valued functions such that
$f(x,0)\to 1$ and $f_x(x,0),\;f_y(x,0)\to 0,\;$ as $|x|\to \infty$,
with the second term in (3.5) vanishing at infinity. The generating
functional of this equation is
$$ F^{(f)}=\int_{-\infty}^{+\infty} dx
\int_0^{+\infty}\left.\left[\frac {1}{2}\ \frac {f_x^2+f_y^2}{f^2-1}+
4(1-f)\right.\right]. $$
The "phase" space ${\cal M}^{(f)}$ is spanned by the functions
$q=\ln |f+\ \sqrt{f^2-1}| $ and $p=f_y/\sqrt{|f^2-1|}, |f|>1$,
and the equations of "motion" are
$$ q_y=\{H^{(f)},q\},\;\;\;  q_{yy}=\{H^{(f)},q_y\},$$
where the "Hamiltonian" is given by

$$
H^{(f)}=\int_{\infty}^{\infty}dx \left.\left[\frac {1}{2}p^2-
\frac {1}{2}\frac {(f_x^2)} {f^2-1}-4(1-f)\right.\right], $$
and the Poisson structure of the "phase" space is defined by the
fundamental brackets $\{p(x),q(x^{\prime})\}=\delta
(x-x^\prime).$\footnote {We stress that it is not our aim here to
construct the Hamiltonian formalism for elliptic equations. In this case,
unfortunately, we do not have "intuitively obvious" variables of the
angle-action type, and all the analogies with the hyperbolic variables
become rather conventional.}

Equation (3.5) is useful because it generates a broad family of
integrable elliptic equations. In what follows, we assume (unless
otherwise stated) that $u=u(z,\bar z)$ is a smooth real-valued function
defined on $R^2_{+}$ and, in addition, that
$$ u(x,y)\to 0, \;\;\; |x| \to 0.\eqno(3.11) $$

Setting $f(z,\bar z)=\cosh u$, we obtain the equation

$$
\triangle u=-4 \sinh u     \eqno(3.12)
$$
from (3.5). Equation (3.12) emerges in the theory of two-dimensional
Boltzmann-Poisson systems at negative temperatures [24] and also in some
reductions of the Chern-Simons model [25]. It has a geometric meaning,
describing the embedding of a negative-curvature surface in
three-dimensional Euclidean space [26].

Using the GE of the problems given by (3.1), (3.2) and by (3.11), (3.12),
which was proved in [4],[5], and also using Eq.(3.4) and the definition
of $f$, we obtain

$$
{\bf S}_x^2=4\cosh ^2 {\frac {u}{2}},\;\;\; {\bf S}_y^2=4\sinh ^2 {\frac
{u}{2}}.\eqno(3.13) $$
These equations, which relate the "potentials" of the two
boundary-value problems, imply, in particular, that the GE does
not impose any constraints on the corresponding "phase" spaces
(provided, of course, that $f>1).$

We list some other possible choices of the functions $f(z,\bar z)$
and the corresponding equations that lead to interesting,
important physical applications,
$$ f(z,\bar z)=-\cosh u,\;\; \partial \bar \partial u=\sinh  u, \eqno(3.14) $$
$$ f(z,\bar z)=e^{\pm u},\;\; \partial \bar \partial u=-2\sinh u \pm \
\frac {\partial u\bar \partial u}{e^{\pm 2u}-1},   \eqno(3.15) $$
$$ f(z,\bar z)=-e^{\pm u},\;\;
\partial \bar \partial u=2\sinh u \pm \ \frac {\partial u\bar \partial
u}{e^{\pm 2u}-1},   \eqno(3.16) $$

$$
f(z,\bar z)=\mp \cos u,\;\; \partial \bar \partial u=\pm \sin u, \eqno(3.17)
$$

$$
f(z,\bar z)=\pm \sin  u,\;\; \partial \bar \partial u=\pm \cos u. \eqno(3.18)
$$

We consider the question of the GE of the problems given by (3.1) and
(3.11), (3.17). For this purpose, we note that Eq. (3.1) can be
rewritten as
$$ c_{ikl}S_k\partial \bar \partial S_l=0, $$
where $c_{ikl}=\epsilon_{ikl},$ with $\epsilon_{ikl}$ being the totally
antisymmetric tensor in $R^3$.
Therefore, Eq. (3.12), as well as Eqs. (3.14)-(3.16)
corresponds to the model of a compact-manifold magnet. This
situation changes for Eqs. (3.16)-(3.18), and the sought for GE is
achieved on a noncompact manifold of matrices $S\in SU(1,1)$.
We set
$$ S= \left(\begin{array}{cc} iS^3& S^1-iS^2 \\ S^1+iS^2& -iS^3
\end{array}\right).\eqno(3.19) $$
The condition $\det S=-1$ and Eq. (3.19) imply that

$$
(S^1)^2+(S^2)^2-(S^3)^2=1,\;\;\; S^2=I.
$$
Geometrically, this means that the end of the vector ${\bf S} = (S^1, S^2,
S^3)$ moves over the surface of the one-sheet hyperboloid (in contrast to
the unit sphere in the "compact" case). The matrix $S$ can be expanded
with respect to the basis elements of the Lie algebra $su(1,1)$ as
$S=S^i\pi_i$, where $\pi_{\alpha}\pi_{\beta}=\eta_{\alpha\beta}+\
ic_{\alpha\beta}^{\gamma}\pi_{\gamma}$, $\alpha, \beta, \gamma=1,2,3,\:
\eta=\{\eta_{\alpha \beta}\}=diag (1,1,-1)$ is the Killing
metric, $c_{\alpha\ \beta}^{\gamma}$ are the structure constants of
$su(1,1)$, and the commutation relations are
$[\pi_\alpha,\pi_\beta]=2ic_{\alpha\beta}^{\gamma}\ \pi_{\gamma}$.

If Eq. (3.17) is now taken with the + sign, it follows from (3.4) and from
the expression for $f$ that
$$
(S_x^1)^2+(S_x^2)^2-(S_x^3)^2=4\sin ^2\frac {u}{2}, $$ $$
(S_y^1)^2+(S_y^2)^2-(S_y^3)^2=-4\cos ^2\frac {u}{2}, $$
and the requirement for conservation of the asymptotic behavior
implies that the GE to (3.1), (3.2) is achieved under the
condition that $u(x,0)\to \pi(mod 2\pi)$. Then, the constraints on the
space $\cal M^{({\bf S})}$ of the corresponding $O(1,2)\;
\sigma$-model take the form of the system of differential inequalities

$$
0\le (S_x^1)^2+(S_x^2)^2-(S_x^3)^2 \le 4,
$$
$$
-4\le (S_y^1)^2+(S_y^2)^2-(S_y^3)^2 \le 0.
$$

The GE conditions for the other equations in the above list are derived in
essentially the same way.

Equations (3.15) and (3.16) seem new. It would be interesting to find
applications of these equations, which could be considered the
Boltzmann-Poisson equations (for positive temperatures in the case of
(3.16) and negative ones in the case of (3.15)) generalized to the case
of a quantum-classical system, and also to quantize them. We
do not rule out other possible applications of these equations
in physics and geometry (a certain regular method to obtain (2
+ 0)-dimensional equations in the differential geometry of surface is
given in [27])\footnote {Exact solutions of Eq. (3.16) were
constructed in [7] using the Darboux transformation method; the
relation of these equations to the Bitsadze equation was also
considered there.}.

The elliptic version of the relativistic field theory model proposed in
[28] takes the following form in the one-component case:

$$
\partial \bar \partial w=-w(1-w^2)-\frac {w\partial w\bar \partial w}{1-w^2},
\;\;\;w=\bar w\eqno(3.20)
$$
Equation (3.20) can be naturally called the elliptic Getmanov
model (a similar equation for a complex-valued function was
considered in [29]). It is known that in the hyperbolic version,
this model is gauge equivalent to the sine-Gordon equation in
Minkowski space [28]. The solution of (3.20) can be easily
related to the solution of Eq. (3.5) as $f=1-2w^2.$ This allows us, in
particular, to obtain from $F^{(f)}$ the density of the generating
functional for (3.20) as
$$\hat F^{(G)}=8(\partial w\ \bar \partial w/(w^2-1)+w^2).$$

Properly speaking, the instances of GE established above between a number
of models on the half-plane and the model described by (3.1) and (3.2)
require a more rigorous (mathematical) proof, which should include the
demonstration of transitions of these models to the $\sigma$-model.
Clearly, this demonstration can be done in each particular case using the
appropriate associated linear problem and constructing the gauge
transformation matrix.

\vskip1cm
\centerline {\bf {4. Exact solutions of the $\sinh$-Gordon equation}}
\vskip1cm

We consider a plasma consisting of electrons and singly charged ions. We
assume that the electron and ion temperatures are the same and each plasma
component has relaxed to the Boltzmann distribution. In this case, the
dimensionless potential of the electric field $u=e\Phi/T$ satisfies the
Boltzmann-Poisson equation [30]

$$
\triangle u=4\sinh  u,\eqno(4.1)
$$
where $ \triangle$ is the two-dimensional Laplace operator written in
dimensionless variables, ${\bf r}={\bf{\bf R}}/2r_D$, $e$ is the electron
charge, $T$ is the temperature, $\Phi$ is the potential,
${\bf R}$ is the radius vector, $r_D=T/(8\pi en_0)^{1/2}$ is the Debye
radius, and $n_0=n_{0e}=n_{0i}$ are the electron and ion concentrations.

Equation (4.1) also emerges in other physical problems: in the
theory of strong electrolytes and conducting films [31], in the
$O(2,1)\: \sigma$-models [32], in the calculation of the electrostatic
contribution to the DNA molecule free energy [33], and so on. The same
equation is obtained by the dimensional reduction (from four to two
dimensions) of the self-duality equations for the $SU(2)$-valued Yang-Mills
fields under a special choice of the ansatz [34]. It was assumed in [34]
that a spontaneous gauge-symmetry breaking occurs: part of the potential
components tend to constant matrices as $r \to \infty$ with half of the
them playing the role of Higgs bosons thereby leading to the massive field
theory corresponding to (4.1).

We also note that there are sufficiently rigorous procedures to derive
(4.1) from the Bogoliubov-Born-Green-Kirkwood-Ivon chain of equations
[35].

We assume that (4.1) is defined on the half-plane $R^2_{+}$,
and that

$$
u(x,y)\to 0 ,\;\;\;  r=\sqrt{x^2+y^2}\to\infty, \eqno(4.2)
$$
where $u=u(x,y)$ is a sufficiently smooth real-valued function.

It is useful to consider first the linear version of (4.1) and (4.2) that
corresponds to the Debye-H\"uckel approximation. In this case, the
solution obtained in the Fourier-transform parametrization is

$$
u(x,y)=\frac {1}{\pi}\int_0^\infty \frac {d\lambda}{\lambda}b_B(\lambda)\
e^{ik(\lambda)x-l(\lambda)y}\;,
$$
$$
\eqno(4.3)
$$
$$
b_B(\lambda)=\frac {1}{4}\int
dxe^{-ik(\lambda)x}\left.\left[l(\lambda)u(x,0)-u_y(x,0)\right.\right], $$
where $k(\lambda)=\lambda-\lambda^{-1}$
and $\l(\lambda)=\lambda+\lambda^{-1}$.  It is assumed that the surface
charge density at the boundary $u_y(x,0)$ decreases faster than
$exp(-2|x|)$ as $|x|\to \infty.$ Then $b_B (\lambda)$ can be analytically
continued from the real axis to the domain in the plane of the complex
variable $\lambda=\lambda_R+i\lambda_I$ that is bounded by the strophoid
curve $\lambda_R^2(2-\lambda_I)-\lambda_I(\lambda_I-1)^2=0$ and that
includes the unit semicircle in the right half-plane. In this domain,
$b_B(\lambda)=b_B(1/{\bar {\lambda}})$ because $u$ is real.

Solution (4.3) occurs under the boundary conditions
$$
b_B(\lambda)=0,\;\;\; \lambda=\bar \lambda< 0,\eqno(4.4)
$$
$$
u(x,0)+\frac {1}{\pi}\int dx^{\prime}      K_0(2|x-x^{\prime}|)u_y(x',0)=0,\
\eqno(4.5)
$$
which follow from the requirement that $u$ be bounded as$r \to \infty$.
In (4.5), $K_0(x)$ is the Macdonald function. Equation (4.5) is the
Fourier transform of (4.4). The asymptotic form of (4.3) as
$r \to \infty$ found using the saddle-point method is
$$ u({\bf r},\alpha)\simeq (\pi r)^{-\frac
{1}{2}}b_B[iexp(-i\alpha)]exp(-2r),\ \eqno(4.6) $$
where $x=r\cos{\alpha}$ and $y=r\sin{\alpha}$. Expression (4.6)
describes the effect of the linear Debye screening.

We return to the original problem formulated in (4.1) and (4.2)
and assume that the given functions $u(x,0)$ and $u_y(x,0)$ cannot be
arbitrary in a given class of functions (for example, in the Schwartz
class) and are constrained by some condition (analogous to (4.5)) to be
formulated in what follows.

The auxiliary linear system of equations corresponding to
(4.1) is
$$ \Psi_x=\left.\left[(i\frac {\lambda}{2}+\frac {\cosh
u}{2i\lambda})\sigma_3- \frac {u_z}{2}\sigma_2-\frac {\sinh
u}{2\lambda}\sigma_1\right.\right]\Psi,\eqno(4.7) $$
$$ \Psi_y=\left.\left[-(\frac
{\lambda}{2}+\frac {\cosh  u}{2\lambda})\sigma_3- i\frac
{u_z}{2}\sigma_2-\frac {\sinh u}{2i\lambda}\sigma_1\right.\right]
\Psi.\eqno(4.8) $$

We introduce the matrix of Jost solutions (4.7), which are determined by
the asymptotic formulas
$$
\Psi^{\pm}=(\Psi_1^{\pm},\Psi_2^{\mp})=e^{ikx\sigma_3/2}(1+o(1)),\;
x\to\pm\infty. \eqno(4.9) $$
We set $$ \Psi^{+}=\Psi^{-}T(\lambda), \;\;
T(\lambda)=\left(\begin{array}{cc} a(\lambda)& b(\lambda)\\ c(\lambda)&
                       d(\lambda) \end{array}\right) ,\; \lambda=\bar
\lambda,\eqno(4.10) $$
where $T(\lambda)$ is the transition matrix. It has the properties
following from (4.7) and (4.10)
$$
T(-\lambda)=\sigma_2T(\lambda)\sigma_2, \;\;\; \bar T(\lambda)=T(\frac {1}{\
\bar \lambda}),\eqno(4.11) $$
and also the unimodularity property $\det  T(\lambda)=1$. From (4.11), we
have

$$
\bar a(\lambda)=a(\frac {1}{\bar {\lambda}}), \;\;\;
\bar b(\lambda)=b(\frac {1}{\bar {\
\lambda}}),\eqno(4.12)
$$
where the second involution takes place, strictly speaking, for
$\lambda=\bar {\ \lambda}$. It, however, admits an analytic continuation
into some domain in the complex plane (we do not need more detailed
information about that domain in what follows). We set
$\phi^{\pm}=\Psi^{\pm}e^{-ikx\sigma_3/2}=(\phi_1^{\pm},\phi_2^{\mp})$.
Then the Volterra integral equations for the direct problem, which are
equivalent to (4.7), become

$$
\phi_j^{\pm}(x,\lambda)=e_j+\int dx^{\prime}g_j^{\pm}(x-x^{\prime},\lambda)\
Q(x^{\prime},\lambda)\phi_j^{\pm}(x^{\prime},\lambda),\eqno(4.13)
$$
where $j=1,2, \; e_1=(1,0)^T,\:e_2=(0,1)^T$, the $Q$ matrix is
$$
Q(x,\lambda)= \left(\begin{array}{cc}
 \frac {\cosh u-1}{2i\lambda}& -\frac {\sinh u}{2\lambda}+\frac {iu_z}{2}\\
-\frac {\sinh  u}{2\lambda}-\frac {iu_z}{2}&
                                   -\frac {\cosh u-1}{2i\lambda}
                       \end{array}\right) ,\;
$$
and $g_j^{\pm}$ are the bare Green's functions
$$
g_1(x,\lambda)=\mp \theta (\mp x)\: diag (1,e^{-ikx}),
$$
$$
\eqno(4.14)
$$
$$
g_2(x,\lambda)=\pm \theta (\pm x)\: diag (e^{ikx},1)
$$
It follows from (4.13) that
$$
\phi_1^{+},\;\phi_2^{-}\in H(\Omega^{+}),\;\;\phi_1^{-},\;\phi_2^{+}\
\in H(\Omega^{-}),\eqno(4.15)
$$
where $H(\Omega)$ is the class of functions that are analytic in the
domain $\Omega$, and $\Omega^{+}=\{\lambda: Im \lambda>0 \}.$

Taking (4.10) and (4.13) into account, we have the integral
representations for the scattering data $a(\lambda)$ ¨ $b(\lambda)$,

$$
a(\lambda)=1-\int dx <e_1^T,\:Q(x,\lambda)\phi_1^{+}(x,\lambda)>,\eqno(4.16)
$$
$$
b(\lambda)=\int dxe^{-ik(\lambda)x}<e_1^T,\:Q(x,\lambda)\phi_2^{+}(x,\
\lambda)>,\eqno(4.17)
$$
where $< , >$ denotes the scalar product of vectors in $C^2$. In the
standard way, Eq. (4.8) implies the relations that describe the spectral
data evolution,

$$
a(y,\lambda)=a(y,0),\;\;\;b(y,\lambda)=b(0,\lambda)e^{-l(\lambda)y},
\eqno(4.18)
$$
from which it can be easily seen that the condition that $b(y,\lambda)$
be finite for any $y>0$ implies
$$ b(\lambda)=0,\;\; \lambda=\bar \lambda<0,\eqno(4.19) $$
and a gap thus appears in the spectrum of the associated linear problem
(see also [3, 36]. Turning to (4.17), we see that a condition similar to
(4.5) emerges in terms of the functions $u(x,0)$ and $u_y(x,0)$.

We consider in more detail the properties of the coefficient $a(\lambda)$.
It follows from (4.16) that $a(\lambda)=1+O(1/\lambda)$ as
$|\lambda|\to\infty$ and $a(\lambda)\in \ H(\Omega^{+})$ and the
unimodularity of $T(\lambda)$ and Eq. (4.19) imply that
$a(\lambda)a(-\lambda)=1$. Therefore, using (4.16), we conclude that
$a(0)=1$. Assuming now that there exist simple zeros of the coefficient
$a(\lambda),  a(\lambda_n)=0 ,\; Im \lambda_n>0,\;n=1,...N$, and recalling
the condition $u=\bar u$ together with Eq. (4.12), we obtain one more
representation for $a(\lambda)$,

$$
a(\lambda)=\prod_{n=1}^{2N_1}\frac {\lambda-\lambda_n}{\lambda+\
 \lambda_n}\prod_{m=2N_1+1}^{2N_1+N_2}\frac {(\lambda-\lambda_m)(\lambda-\
\frac {1}{\bar \lambda_m})}{(\lambda+\lambda_m)(\lambda+\
\frac {1}{\bar \lambda_m})},\eqno(4.20)
$$
where $2N_1+2N_2=N$ and the zeros $\lambda_n,\; 1\le n\le 2N_1,$ belong to
the unit circle (the analogues of kinks for the hyperbolic
version of the $\sin$-Gordon equation [21]). The zeros
$\lambda_m$ and $1/{\bar \lambda_m},\; 2N_1+1\le m\le 2N_1+N_2$
constitute an inversion with respect to the unit circle (the analogues of
breathers [21]). It is obvious that there is no topological charge in the
model, in contrast to the equation $\triangle u=\sin u$ (see [37, 38]).

It follows from the first relation in (4.18) that $\ln  a(\lambda)$,
in particular, can be considered a generating functional of the integrals
of "motion". The standard procedure brings the system of equations for the
elements of the column $\phi_1^{+}$ to the Riccati equation

$$
F_x+Q_{12}F^2=-ik(\lambda)F+(Q_{22}-Q_{11})F+Q_{21},\eqno(4.21)
$$
where $F=\phi_{21}^{+}/\phi_{11}^{+}$. Further, we obtain

$$
\ln  a(\lambda)= -\int_{\infty}^{\infty} dx[Q_{11}+Q_{12}F].\eqno(4.22)
$$
Setting

$$
F=\sum_{k=0}^{\infty}F_{-k}\lambda^k,\;|\lambda|\to 0,\;\;
F=\sum_{k=1}^{\infty}F_{k}\lambda^{-k},\;|\lambda|\to \infty,
$$
$$
R=Q_{11}+Q_{12}F=\sum_{k=0}^{\infty}R_{-k}\lambda^{k},\;\
|\lambda|\to 0,\;\;
R=\sum_{k=1}^{\infty}R_{k}\lambda^{-k},\;|\lambda|\to \infty,
$$
we see from (4.21) that
$$
R_{-1}=i[(\cosh u-1)/2+u_z^2/2]+iu_x^2/2-\
i\partial_x^2\ln (\cosh u/2)-(\cosh u-1)/2\partial_x(u_z/\sinh  u),
$$
$$
R_0=\partial_x\ln \cosh u/2,\;\: R_1=-i[(\cosh  u-1)/2+u_z^2/2),\;\:
R_2=-(1/8)\partial_x(u_z^2),
$$
and so on. Since the left-hand side of (4.22) does not depend on
$y$, the rigth-hand side can be evaluated at
$y=0$, which gives an infinite number of conservation laws.

Now let

$$\ln a(\lambda)=\sum_{s=1}^{\infty}I_s\lambda^{-s},\;
|\lambda|\to {\infty} $$
and
$$\ln a(\lambda)=\sum_{s=0}^{\infty}\
I_{-s}\lambda^s,\; |\lambda|\to 0.$$
It follows from (4.12) that $I_s=\bar I_{-s},\;I_{2s}=0$ and $I_0=0$.
Using (4.20)-(4.22), we arrive at the trace identity

$$\frac {2}{2s-1}[\sum_{m=1}^{2N1}\lambda_m^{2s-1}+\sum_{n=2N_1+1}^
{2N_1+N_2}(\
\lambda_n^{2s-1}+\bar \lambda_n^{-2s+1})]=\int dxR_{2s-1}(x),\eqno(4.23)
$$
where $s = 1,2,....$ The functions $R_{2s}$ are either equal
to zero or given by total derivatives in $x$, their integrals vanishing in
view of (4.2).

We now solve the inverse problem. The standard argument employing
the analytic properties of solutions $\phi_{1,2}^{\pm}$ leads to
the Riemann problem of reconstructing a piecewise-analytic
vector function from its jump on the boundary
$(\lambda=\bar \lambda)$:
$$ \frac {\phi_2^{+}(x,y,\lambda)}{a(\lambda)}=\phi_2^{-}(x,y,\lambda)+\
r(y,\lambda) e^{ik(\lambda)x}\phi_1^{+}(x,y,\lambda),\eqno(4.24)
$$
where $r(y,\lambda)=b(y,\lambda)/a(\lambda)=b(0,\lambda)/a(\lambda)
e^{-l(\lambda)y}=r(0,\lambda)e^{-l(\lambda)y}$
is the reflection coefficient. After some calculations, we finally obtain
the system of singular integral inverse scattering equations
$$
\phi_1^{+}=e_1+i\sum_{n=1}^{N}\frac {m_n(y)e^{ik(\lambda_n)x}}{\lambda+\
\lambda_n}\sigma_2\phi_{1n}^{+}(x,y)-\int_0^{\infty} \frac {d\mu}{2\pi}\
\frac {r(y,\mu)e^{ik(\mu)x}}{\mu+\lambda+i0}\sigma_2\phi_1^{+}(\mu),\
\eqno(4.25)
$$
$$
\phi_{1m}^{+}=e_1+i\sum_{n=1}^{N}\frac {m_n(y)e^{ik(\lambda_n)x}}{\lambda_m+\
\lambda_n}\sigma_2\phi_{1n}^{+}(x,y)-\int_0^{\infty} \frac {d\mu}{2\pi}\
\frac {r(y,\mu)e^{ik(\mu)x}}{\mu+\lambda_m}\sigma_2\phi_1^{+}(\mu),
\eqno(4.26)
$$
where $N=2(N_1+N_2)$ and $m_n=b_n/a^{\prime} (\lambda_n)$ are the discrete
spectrum transition coefficients such that $m_n=-\lambda_n^2\bar m_n$ for
$1\le n \le 2N_1$ and $m_n=-\bar \lambda_n^2m_{n+N_2}$ for $2N_1+1\le n
\le 2N_1+N_2.$ In Eqs. (4.25) and (4.26), further,
$\phi_1^{+}(x,y,\lambda),\;$ and $\phi_{1n}^{+}(x,y)$ are the respective
eigenfunctions of the continuous and the discrete spectra.

The formulas for reconstructing the potential should be added to
the system given by (4.25) and (4.26). Setting
$$ \phi^{+}(x,y,\lambda)=\sum_{k=0}^\infty
\phi_k^{+}(x,y)\lambda^k,\;\;\;|\lambda|\to 0, $$
we insert this expansion into the matrix equation as $\phi^{+}$
and compare it with the corresponding expansion (4.7). We thus obtain one
of the formulas for reconstructing the potential,

$$ i\sinh  \frac {u}{2}=\sum_{n=1}^{N}\frac
{\phi_{11n}^{+}e^{ik(\lambda_n)x}}{\ \lambda_n}m_n(y)-\int_0^{\infty}\frac
{d\mu}{2\pi i}\frac {r(y,\mu)}{\mu}\
\phi_{11}^{+}(x,y,\mu)e^{ik(\mu)x}.\eqno(4.27)
$$

A similar expansion for $|\lambda| \to \infty$ yields
$$
\frac {u_x-iu_y}{4}=\sum_{i=1}^{N} \phi_{11n}^{+}e^{ik(\lambda_n)x}m_n(y)-\
\int_0^{\infty} \frac {d\mu}{2\pi i}r(y,\mu)\phi_{11}^{+}(x,y,\mu)
e^{ik(\mu)x}.\eqno(4.28)
$$

Relations (4.25)-(4.28) allow us to construct the simplest solutions of
Eq. (4.1). We first consider the "no-soliton" case $a(\lambda)=1$.
Then system (4.25), in which $r \to \infty, $ can be easily
iterated, the first iteration giving $\phi_{11}^{+}(r,\lambda)=1+O(1/r).$
Using (4.27) and also the fact that $u(r,\alpha) \to 0$ as $r \to \infty$
for $\alpha \in [0,\pi],$ we obtain
$$ u(r,\alpha) \cong \int_0^\infty
\frac {d\mu}{\pi}\frac {b(0,\mu)}{\mu}\ e^{rf(\mu)},\eqno(4.29) $$
where $f(\mu)=ik(\mu)\cos \alpha-l(\mu)\sin \alpha$. Assuming that
$b(\mu)$ can be analytically continued into the required domain in the
complex plane, we obtain

$$
u(r,\alpha) \cong \frac {b(0,ie^{-i\alpha})}{\sqrt {\pi\
r}}e^{-2r}[1+O(\frac {\ e^{-2r}}{r})],\;\; r \to \infty \eqno(4.30) $$
Comparing this with (4.6), we see that the only difference in the
two expressions is the replacement $b_B\to b$ and that (4.30) becomes
(4.6) with sufficiently weak fields. This effect can be called a
quasi-linear Debye screening: the scale of the screening is the same as in
the linear case, and the influence of strong fields amounts to
renormalization of the preexponential factor - the effective charge of the
boundary - which contributes to the asymptotic behavior of the potential.

We now set $b(\lambda)=0,\;0 \le \lambda \le \infty,$ which means that
we consider the case of reflectionless potentials. The system of
inverse scattering transform equations then reduces to a system of
linear algebraic equations. Recalling (4.28), we obtain the
"$N$-soliton" solution

$$
u(x,y)=2\ln \frac {\Delta^{+}}{\Delta^{-}},\;\;
\Delta^{\pm}=\det ({\delta_{mn}\pm \frac {im_n(y)}{\lambda_m+\lambda_n}\
\exp {i{\frac {x}{2}[k(\lambda_n)-k(\lambda_m)]}}}).\eqno(4.31)
$$
where $1 \le m,n \le N$. One of the simplest solutions (the analogue
of a breather) corresponds to setting $N_1=0,\;N_2=1,\;\lambda_1=1/\bar
\lambda_2=\exp(\gamma+i\theta),$
and $m_1(0)=\lambda_1^2m_2(0)=\exp(\delta+i\rho )$. It follows from (4.31)
that

$$
u(x,y)=2\ln \frac {1+h}{1-h},\;\;
h=\frac {\sin  [2r\sinh  \gamma \cos  (\theta+\alpha)+\rho -\theta)]\coth
\gamma}{\ \cosh  [2r\cosh  \gamma \sin  (\theta+\alpha)+\gamma-\delta-\ln
\frac {\tanh \gamma}{2}]}. \eqno(4.32) $$
In the general case where
$(\theta+\alpha \ne 0,\pi),$ this solution describes a charge-density
wave, which falls off as $r \to \infty$ according to the law
$$ u \sim \exp (-2r\cosh  \gamma |\sin (\theta+\alpha)|.\eqno(4.33)
$$
The parameters involved in (4.32) can be related to the values
of the field at the boundary $y=0$ by the trace identities
(4.23), which now become

$$
\cosh \gamma \cos \theta =-\frac {1}{32}\int dx u_x(x,0)u_y(x,0),\eqno(4.34)
$$
$$
\cosh {\gamma}{\sin \theta}=-\frac {1}{32} \int dx [4(\cosh u(x,0)-1)+\
\frac {1}{2}(u_x^2(x,0)-u_y^2(x,0))].\eqno(4.35)
$$
As can be seen from (4.32), the field does not fall off for
$\theta+\alpha=\pi$, which means that a long-range order occurs in that
direction, i.e., there is a coherent structure describing the
distribution of the electric field and plasma component densities with a
considerable charge separation. We also note that for
$h=\pm 1$, solution (4.32) becomes singular (we do not consider this case
here).

Another simple solution (the analogue of a kink) that follows from (4.31)
is obtained by taking
$N_1=1,\;N_2=0,\;\lambda_n=e^{i\theta_n},n=1,2,\;m_{1,2}(0)=i\
e^{\delta_{1,2}+i\theta_{1,2}}$:
$$ h=\frac {\cosh [2r\sin (\frac
{\theta_1-\theta_2}{2})\cos (\frac {\theta_1+\theta_2}{\ 2}+\alpha)-\frac
{\delta_1-\delta_2}{2}]\cot \frac {\theta_1+\theta_2}{2}}{\sinh [2r\cos
(\frac {\theta_1-\theta_2}{2})\sin (\frac {\theta_1+\theta_2}{2}+\
\alpha)-\frac {\delta_1+\delta_2}{2}-\ln (\frac {1}{2}\tan \frac {\theta_1+\
\theta_2}{2})]}.\eqno(4.36) $$
This case corresponds to the nonlinear Debye screening. As
$r \to \infty$ , we have

$$
h \sim e^{-2r\sin (\theta_1+\alpha)+2\delta_1}+e^{-2r\sin (\theta_2+\alpha)+\
2\delta_2},\; 0<\frac {\theta_1+\theta_2}{2}+\alpha<\pi.\eqno(4.37)
$$
The term with the smallest coefficient in front of $2r$ survives in the
exponent of (4.37). Because this coefficient is always less than one, the
field falls off more slowly under the nonlinear Debye screening than in
the quasi-linear case, and, moreover, the fall-off is anisotropic. The
parameters characterizing the fall-off (4.37) can be related to the
boundary values of the fields using trace formulas (4.34) and (4.35) if we
replace $\cosh \gamma$ there by  $\cos \frac {\theta_1-\theta_2}{2}$, and
also $\theta$ by $\frac {\theta_1+\theta_2}{2}$.

The slower fall-off of the potential under the nonlinear Debye screening
can be explained by the pressure of the electric field, which "pushes
aside" the electrons and ions in the polarization cloud surrounding the
boundary. The collective nonlinear phenomena observed here - the
threadlike "radiographic examination" of the plasma with the electric
field and its slower fall-off law - seem to take place only in the
two-dimensional case. The field between the piontlike charges then
decreases more slowly than in three dimensions, which explains the
appearance of long-range correlations.

We now consider several thermodynamic relations and also those involving
energy. The full energy of the medium (using dimensionless quantities) is

$$
E=\int dxdy [(\nabla u)^2+4u\sinh  u].\eqno(4.38)
$$
In the linear case, this relation can be represent as

$$
E=\int dxdy [(\nabla u)^2+4u^2]=E_f^{(x)}+E_f^{(y)}+E_p.\eqno(4.39)
$$
Using (4.3) and the relation $$\int dx
e^{i[k(\lambda)+k(\mu)]x}=(2\pi/(\lambda\
l(\lambda)))\delta(\mu-1/\lambda),$$ we evaluate
$E_f^{(x)},E_f^{(y)}$ and $E_p$ with the result

$$
E_f^{(x)}=\frac {1}{\pi}\int_0^{\infty} \frac {d\lambda}{\lambda}|\
b_B(\lambda)\
|^2(\frac {k(\lambda)}{l(\lambda)})^2,\;\;\;
E_f^{(y)}=\frac {1}{\pi}\int_0^{\infty} \frac {d\lambda}{\lambda}
|b_B(\lambda)|^2,\eqno(4.40)
$$
where
$$
E=\frac {2}{\pi}\int_0^{\infty} \frac {d\lambda}{\lambda}|b_B(\lambda)|^2.\
\eqno(4.41)
$$
Restricting ourselves to the "no-soliton" sector in the nonlinear
case, we can exactly evaluate the quantities analogous to those
found above only for the sum $E_f^{(x)}+E_f^{(y)}$ and only for the
continuous spectrum (the integrals diverge in the other cases). From
(4.28), we see that

$$
\frac {u_x-iu_y}{4}=-\int_0^{\infty} \frac {d\mu}{2\pi i\mu}r(y,\mu)\
\chi_{11}^{+}(x,y,\mu),
$$
where we introduce the function
$\chi_1^{\pm}(x,y,\lambda)=\phi_1^{\pm}(x,y,\lambda)e^{ik(\lambda)x}.$
It now follows that
$$
\int dxdy (u_x^2+u_y^2)=4\int_0^{\infty}\frac {d\mu d\lambda}{\pi \mu      \
\lambda}r(0,\mu) \bar r(0,\lambda)\int_0^{\infty} dye^{-[l(\lambda)+\
l(\mu )]x}\int dx \chi_{11}^{+}(x,\mu)\bar \chi_{11}^{+}(x,\lambda).\
\eqno(4.42)
$$
To evaluate the integral, therefore, we must know the orthogonality
relations for the functions $\chi_{11}^{+}(x,\lambda).$ The method for
obtaining these is to write the integral equations for
$\chi_{11}^{+}(x,\lambda)$ and $\bar \chi_{11}^{+}(x,\lambda)$ and then
multiply them using the formulas
$$ \int dx  \theta(\xi-x)e^{-i[k(\lambda)-k(\mu)](x-\xi)}=\frac {\pi \mu
\delta (\ \lambda-\mu)}{l(\mu )}-iP\frac {1}{k(\lambda)-k(\mu)}, $$
$$ \int dx \theta(\xi_1-x)\theta(\xi_2-x)\
e^{-i[k(\lambda)-k(\mu)]x-ik(\lambda)\xi_1+ik(\mu)\xi_2}=(\
\frac {\pi \mu \delta (\
\lambda-\mu)}{l(\mu )}-iP\frac {1}{k(\lambda)-k(\mu)})\Gamma_0,
$$
where $P$ denotes the principal value and $\Gamma_0=e^{-ik(\mu)\
\xi_1+ik(\mu)\
\xi_2}$,
$\xi_1>\xi_2$; $e^{ik(\lambda)\xi_2-ik(\lambda)\xi_1}$, $\xi_2>\xi_1\;\
;1$, $\xi_1=\xi_2.$ After some calculations, we find

$$
\int dx \chi_{11}^{+}(x,\mu )\bar \chi_{11}^{+}(x,\lambda)=\
\frac {\pi \mu \delta (\lambda-\
\mu)}{l(\mu )}(|a(\mu)|^2+1)-iP\frac {a(\lambda)\bar a(\mu)-1}{k(\lambda)-\
k(\mu)},
$$
and finally

$$
\int dx dy (u_x^2+u_y^2)=\frac {1}{2\pi}\int_{0}^\infty \frac {d\mu}{\mu}\
|b(\mu)|^2\left.\left(1+\frac {1}{|a(\mu)|^2}\right.\right)
\left.\left(\frac {k^2(\mu)}{l^2(\mu)}+1\right.\right).\eqno(4.43)
$$
In the linear limit as $u \to 0
\;(a(\mu)=1)$ and $b(\mu) \to b_B(\mu),$ this expression becomes equal to
(4.41).

We consider the problem of the charge of the medium (as a whole). Its
total value is

$$
4\int_{R^2} dx dy \sinh u=\int_{R^2} dx dy \triangle u=
-\int d\xi u_y(\xi,0).\eqno(4.44)
$$
$$\empty$$
Whenever the solution $u$ is singular, a term given by the sum
of integrals along the contour that cuts out the singularities
should be added to the right-hand side of (4.44). Using the
"breather" solution and evaluating this sum, we can show that
this term vanishes and we have (4.44) again. This relation,
therefore, shows that the electric neutrality condition should
be understood such that the total charge of the medium and of
the boundary vanishes.

The trace identities (4.34) and (4.35) also have a very clear
physical interpretation. To show this, we introduce the Maxwell
stress tensor of the electric field:
$T_{\alpha \beta}=\hat E_{\alpha}\hat E_{\gamma}-\ (1/2)\delta_{\alpha
\gamma}\hat{\bf E}^2$, where $\hat {\bf E}=(\hat E_x,\hat E_y)$ is the
electric field vector. Because the field has a potential, we obtain

$$
[curl \hat {\bf E}\:,\:\hat {\bf E}]_{\alpha}=\partial_{\gamma}T_{\alpha
\gamma}-2T_0\ \partial_{\alpha}(\cosh  \frac {e\Phi}{T}-1),\eqno(4.45) $$
where
$$
T_{\xi \eta}=\Phi_{\xi}\Phi_{\eta},\;\; T_{\eta \eta}=-\frac {1}{2}(\
\Phi_{\xi}^2-\Phi_{\eta}^2),\eqno(4.46)
$$
$\xi=(\sqrt T/\alpha )x,\;\eta=(\sqrt T/\alpha )y$ and $\alpha=\sqrt{2\pi
e^2n_0}.$ It follows from (4.45) that

$$
\partial_{\eta}\left[ {T_{\eta \eta}-2n_0T(\cosh \frac
{e\Phi}{T}-1)}\right] =-\partial_{\xi} T_{\eta \xi}.\eqno(4.47) $$
On the other hand, evaluating the force applied to the boundary from
the medium, we obtain $\hat F_{\xi}={ 0}$, and
$$ \hat  F_{\eta}=2en_0\int_{R_{+}^2} d\xi d\eta \sinh \frac
{e\Phi}{T}\Phi_{\eta}= -2Tn_0\int d\xi (\cosh  \frac
{e\Phi}{T}-1).\eqno(4.48) $$
Comparing (4.35)-(4.48) with (4.34) and (4.35), we see that the "trace
identities" involve the quantities that can be expressed through the
Maxwell stress tensor and the force applied to the boundary.

We consider several properties of the parameter $\gamma$, entering
(4.34) and (4.35), where we assume that $\gamma=\gamma(T,n_0)$. We note
that the "breather" solution degenerates as $\gamma \to 0$, all the
eigenvalues of the auxiliary linear problem becoming equal. This
degeneration occur in two ways: ( ) at $n_0=n_c$, where $T$ is arbitrary;
(b) at $T=T_c$ where $n_0$ is arbitrary. Expanding the right-hand sides of
these identities in the power series in
$(n-n_c)$ or $(T-T_c)$, we find the behavior of $\gamma$ near the
critical parameters $n_c$ and $T_c$.

In case a, it follows from (4.34) that
$\gamma \sim \sqrt{2(n_c-n)/n_c},$ as $n \nearrow n_c.$
This behavior of $\gamma$ occurs under the additional
condition, which follows from (4.35),

$$
\int d\xi (\cosh \frac {e\Phi_c}{T}-1)=0,\;\;
\Phi_c=\Phi(\sqrt{\frac {2\pi e^2\
n_c}{T}}\xi,0).\eqno(4.49)
$$
A similar analysis in case b results in $\gamma \sim \sqrt{2(T_c-T)/T}$ as
$T \nearrow T_c$, with condition (4.49) replaced by

$$
4\int d\xi (\cosh  \frac {e\Phi_c}{T_c}-1)=\int d\xi \frac {e\Phi_c}{T_c}\
\sinh \frac {e\Phi_c}{T_c},\
\;\; \Phi_c=\Phi (\frac {\alpha \xi}{\sqrt {T_c}},0).\eqno(4.50)
$$

The degeneracy of the eigenvalues can probably be interpreted in this case
as a certain "phase" transition in the nonlinear medium. This hypothesis,
however, requires a further analysis.

Note also, that the equation

$$
\triangle u=-4 \sinh u,  \eqno(4.51)
$$
describing the plasmas at the negative temperatures, was solved in [39].
\vskip0.4cm
\centerline{\bf{Acknowlegements}}
\vskip0.3cm

The authors are grateful to A.G. Izergin, P.P. Kulish, and V.N.
Krasil'nikov for their attention to this work and their support at various
stages.

This work was supported by Russian Foundation for Basic Research (Grants
NNo. 96-01-00548, 98-01-01063).
\newpage

\centerline{\bf {REFERENCES}}
\vskip2cm
[1]  \parbox[t]{12.7cm}
     {{\em V.D.Lipovskii and S.S.Nikulichev, }
       Vestn. Leningr. Gos. Univ., Ser. Fiz., Khim., No. 4 (25), 61 (1989).}
\vskip0.3cm
[2]  \parbox[t]{12.7cm}
     {{\em E.Sh.Gutshabash and V.D.Lipovskii}, Zap. Nauchn. Sem. LOMI,
          {\bf 180}, 53(1990).}
\vskip0.3cm

[3] \parbox[t]{12.7cm}
     {{\em E.Sh.Gutshabash and V.D.Lipovskii}, Theor. Math. Phys.,
{\bf 90}, 175 (1992).}
\vskip0.3cm

[4]  \parbox[t]{12.7cm}
     {{\em G.G.Varzugin, E.Sh.Gutshabash and V.D.Lipovskii
}, Theor. Math. Phys., {\bf 104}, 513 (1995).}
\vskip0.3cm

[5]  \parbox[t]{12.7cm}
     {{\em E.Sh.Gutshabash}, Vest. Leningr. Gos. Univ., Ser. Fiz., Khim.,
      No.4 (25), 84 (1990).}

\vskip0.3cm

[6] \parbox[t]{12.7cm}
     {{\em   E.Sh.Gutshabash and V.D.Lipovskii}, Zap. Nauchn. Sem.
            LOMI, {\bf 199}, 71 (1992).}
\vskip0.3cm

[7] \parbox[t]{12.7cm}
     {{\em E.Sh.Gutshabash}, Zap. Nauchn. Sem. LOMI, {\bf 245}, 149 (1997).}
\vskip0.3cm

[8] \parbox[t]{12.7cm}
    {{\em A.V.Belinskii and V.E.Zakharov}, Journal eksperimentalnoi i
      teoreticheskoi fiziki, {\bf 75}, 1953 (1978).}
\vskip0.3cm

[9] \parbox[t]{12.7cm}
    {{\em A.V.Belinskii and V.E.Zakharov}, Journal eksperimentalnoi i
      teoreticheskoi fiziki, {\bf 77}, 1 (1979).}
\vskip0.3cm

[10] \parbox[t]{12.7cm}
    {{\em D.Maison}, Phys.Rev.Lett., {\bf 41}, 521 (1978).}
\vskip0.3cm

[11] \parbox[t]{12.7cm}
    {{\em D.Maison}, J.Math.Phys., {\bf 20}, 871 (1979).}
\vskip0.3cm

[12] \parbox[t]{12.7cm}
   {{\em  F.A.Bais and R.Sasaki}, Nucl.Phys.B. {\bf 202}, 522 (1982).}

\vskip0.3cm

[13] \parbox[t]{12.7cm}
  {{\em  I.Hauser and F.J.Ernst}, J.Math.Phys.,{\bf 22}, 1051 (1981).}
\vskip0.3cm

[14] \parbox[t]{12.7cm}
     {{\em G.G.Varzugin}, Teor.Mat.Phys ,
      {\bf 111}, 345 (1997).}
\vskip0.3cm

[15] \parbox[t]{12.7cm}
    {{\em A.V.Mikhailov and A.I.Yaremchuk}, Nucl.Phys.B.,
     {\bf 202}, 508 (1982).}

\vskip0.3cm

[16] \parbox[t]{12.7cm}
   {{\em  A.V.Mikhailov }. In: "Solitons. Modern problems in condensed
      matter. Ed.by S.E.Trullinger , V.L.Pokrovskii and V.E. Zakharov,
  North. Holl. Publ. Co., {\bf 17}, 1986.}
\vskip0.3cm

[17] \parbox[t]{12.7cm}
   {{\em P.Baxter}. Exactly Solved Models in Statistical Mechanics,
     Acad. Press, London (1982)}

\vskip0.3cm

[18] \parbox[t]{12.7cm}
  {{\em Yu.A.Izyumov and Yu.N.Skriabin}, Statistical Mechanics of
    Magnetically Ordered Systems, Plenum Press, New York (1988).}

\vskip0.3cm

[19] \parbox[t]{12.7cm}
  {{\em E.M.Lifschitz and L.P.Pitaevski}, Electrodynamics of Continuos
    Substances, Nauka, Moscow (1982) (in Russian).}
\vskip0.3cm

[20] \parbox[t]{12.7cm}
    {{\em  S.P.Burtsev, V.E.Zakharov and A.V.Mikhailov}, Teor.Math.Phys.
    {\bf 70}, 323 (1987).}
\vskip0.3cm

[21] \parbox[t]{12.cm}
  {{\em L.D.Faddeev and L.A.Takhtadzhyan}, The Hamiltonian Methods in the
    Theory of Solitons , Springer, Berlin (1987).}
\vskip0.3cm

[22] \parbox[t]{12.7cm}
  {{\em Yu.A.Izyumov}, Neutron Diffraction in Long-Periodic Structures,
   Energoatomizdat, Moscow (1987).}
\vskip0.3cm

[23] \parbox[t]{12.7cm}
  {{\em K.Pohlmeyer }, Commun. Math. Phys., {\bf 46}, 207 (1976).}
\vskip0.3cm

[24] \parbox[t]{12.7cm}
    {{\em G.Joyce and D.Montgomeri }, J.Plasma Phys. {\bf 10}, 107 (1973).}
\vskip0.3cm

[25] \parbox[t]{12.7cm}
    {{\em R.Jackiw and So-Young-Pi}, Prog. Theor. Phys. Suppl. {\bf 107},
     (1992).}
\vskip0.3cm

[26]  \parbox[t]{12.7cm}
    {{\em  A.I.Bobenko}, Russ. Math. Surv., {\bf 46}, 1 (1991).}
\vskip0.3cm

[27]  \parbox[t]{12.7cm}
    {{\em   E.G.Poznyak and A.G.Popov}, Russ. Acad. Sci., Dokl., Math.,
	    {\bf 48}, 338 (1994).}
\vskip0.3cm

[28]  \parbox[t]{12.7cm}
   {{\em  B.S.Getmanov}, JETP Lett., {\bf 25}, 119 (1977).}
\vskip0.3cm

[29]  \parbox[t]{12.7cm}
   {{\em  I.V.Barashenkov and B.S.Getmanov}, J. Math. Phys.
       , {\bf 34}, 3054 (1993).}
\vskip0.3cm

[30]  \parbox[t]{12.7cm}
   {{\em  G.Ecker}, Theory of Fully Ionized Plasmas, Acad. Press, New York
       (1972).}
\vskip0.3cm

[31]  \parbox[t]{12.7cm}
  {{\em   S.Yu.Dobrokhotov and V.P.Maslov}, Zap. Nauchn. Sem. LOMI,
       {\bf 84}, 35 (1979).}
\vskip0.3cm

[32]   \parbox[t]{12.7cm}
   {{\em  V.E.Chelnokov and M.G.Tseitlin }, Phys.Lett.A.,
      {\bf 99}, 147 (1983).}

\vskip0.3cm

[33]   \parbox[t]{12.7cm}
  {{\em  M.D.Frank-Kamenetskii, V.D.Anshelevich and A.V.Lukashin
}, Sov. Phys. Usp., {\bf 151}, 317 (1987).}

\vskip0.3cm

[34]   \parbox[t]{12.7cm}
  {{\em  Ch.K.Saclioglu}, J. Mat. Phys., {\bf 25}, 3214 (1984)}

\vskip0.3cm

[35] \parbox[t]{12.7cm}
       {{\em D.Montgomery and G.Joyce }, Phys. Fluids,
{\bf 17}, 1139 (1974)}
\vskip0.3cm

[36] \parbox[t]{12.7cm}
{{\em M.Jaworski and D.Kaup }, Preprint Ins. for
Studies. Potsdam: INS, 1989.}
\vskip0.3cm

[37] \parbox[t]{12.7cm}
{{\em A.B.Borisov}, "Nonlinear exitations and two-dimensional
topological solitons in magnets", Doctorial dissertation, Inst. Fiz. Metallov,
Sverdlovsk, 1986.}
\vskip0.3cm

[38] \parbox[t]{12.7cm}
{{\em A.B.Borisov A.B. and V.V.Kiselev }, Physica D.,
{\bf 31}, 49 (1988).}
\vskip0.3cm

[39] \parbox[t]{12.7cm}
{{\em E.Sh.Gutshabash}, Zap.Nauch.Sem.LOMI, {\bf 251}, 251(1998).}

\end{document}